\documentclass[aps,pre,reprint,showpacs,superscriptaddress]{revtex4-1}
\usepackage{graphicx}
\usepackage{dcolumn}
\usepackage{bm}
\usepackage{color}
\usepackage{lipsum}
\usepackage{mathrsfs}
\usepackage{lineno}
\usepackage{siunitx}
\usepackage{dcolumn}
\usepackage{amsmath}
\usepackage{hyperref}
\hypersetup{hypertex=true,colorlinks=true,linkcolor=blue,anchorcolor=blue,citecolor=blue,urlcolor=blue}
\usepackage{enumerate}
\usepackage{multirow}
\usepackage{appendix}
\usepackage{subfigure}
\begin{document}
\title{Assessing Proton-Boron Fusion Feasibility under non-Thermal Equilibrium Conditions: Rider's Inhibition Revisited}
\author{S. J. Liu}
\affiliation{Institute for Fusion Theory and Simulation, School of Physics, Zhejiang University, Hangzhou, 310058, China}
\author{D. Wu}
\email{dwu.phys@sjtu.edu.cn}
\affiliation{Key Laboratory for Laser Plasmas and School of Physics and Astronomy, and Collaborative Innovation Center of IFSA (CICIFSA), Shanghai Jiao Tong University, Shanghai, 200240, China}
\author{B. Liu}
\email{liubing@enn.cn}
\affiliation{Hebei Key Laboratory of Compact Fusion, and ENN Science and Technology Development Co., Ltd, Langfang 065001, China}
\author{Y.-K.M. Peng}
\affiliation{Hebei Key Laboratory of Compact Fusion, and ENN Science and Technology Development Co., Ltd, Langfang 065001, China}
\author{J. Q. Dong}
\affiliation{Hebei Key Laboratory of Compact Fusion, and ENN Science and Technology Development Co., Ltd, Langfang 065001, China}
\author{T. Y. Liang}
\affiliation{Institute for Fusion Theory and Simulation, School of Physics, Zhejiang University, Hangzhou, 310058, China}
\author{Z. M. Sheng}
\affiliation{Institute for Fusion Theory and Simulation, School of Physics, Zhejiang University, Hangzhou, 310058, China}
\date{\today}


\begin{abstract}
Compared to the D-T reaction, the neutron-free proton-boron (p-$^{11}$B) fusion has garnered increasing attention in recent years. However, significant Bremsstrahlung losses pose a formidable challenge in p-$^{11}$B plasmas in achieving $Q>1$ in thermal equilibrium.
The primary aim of this study is to corroborate Todd H. Rider's seminal work in the 1997 Physics of Plasmas, who investigated the feasibility of sustaining p-$^{11}$B fusion under non-thermal equilibrium conditions.
Employing a series of simulations with new fusion cross-section, we assessed the minimum recirculating power that must be recycled to maintain the system's non-thermal equilibrium and found that it is substantially greater than the fusion power output, aligning with Rider's conclusions, whether under the conditions of non-Maxwellian electron distribution or Maxwellian electron distribution, reactors reliant on non-equilibrium plasmas for p-$^{11}$B fusion are unlikely to achieve net power production without the aid of highly efficient external heat engines.
However, maintaining the ion temperature at 300 keV and the Coulomb logarithm at 15, while increasing the electron temperature beyond 23.33 keV set by Rider, leads to diminished electron-ion energy transfer and heightened Bremsstrahlung radiation.
When the electron temperature approaches approximately 140 keV, this progression ultimately leads to a scenario where the power of Bremsstrahlung loss equals the power of electron-ion interactions, yet remains inferior to the fusion power. Consequently, this results in a net gain in energy production.
\end{abstract}

\maketitle

\section{Introduction}
In recent years, researches on the p-$^{11}$B fusion have made significant progress both in thermonuclear reactions and beam-target fusion \cite{Magee2023, Margarone2022, Liusj2024}.
The p-$^{11}$B fusion holds great advantages over the D-T reaction due to its minimal production of neutrons.
Unlike other neutronless reactions, which either suffer from material scarcity (such as the D-$^{3}$He reaction), excessive neutron production in secondary reactions (like the D-D reaction), or possess lower cross-sections than p-$^{11}$B fusion (such as the p-$^{6}$Li reaction), p-$^{11}$B fusion stands out.
However, achieving the maximum reaction cross-section for p-$^{11}$B fusion requires a significantly higher center-of-mass energy compared to the D-T reaction. This necessitates high ion temperatures to achieve the requisite reactivity for thermal fusion systems \cite{Atzeni2004}.

Meanwhile, the electron temperatures also escalate significantly, inducing intense Bremsstrahlung radiation, which poses a challenge to convert into utilizable energy.
Previous analyses have indicated that radiation losses attributed to Bremsstrahlung are overwhelmingly high, thereby impeding the achievement of a burning plasma in equilibrium conditions, ultimately yielding no net energy output \cite{Moreau1977, Nevins1998, Nevins2000}.

Todd H. Rider used to conduct fundamental analyses on plasma fusion systems operating outside of thermodynamic equilibrium \cite{Rider1995phd, Rider1997}. Rider proposed that a minimum power must be recycled to sustain the non-thermal equilibrium state despite collisions, termed as the $recirculating\ power$.
Utilizing the Fokker-Planck equation, he determined the minimum recirculating power. Furthermore, He calculated the power of nuclear fusion and  Bremsstrahlung loss. 
Rider's analyses revealed that when Bremsstrahlung power reaches approximately half of the fusion power, the recirculating power consistently surpasses the fusion power significantly, irrespective of whether the electron distribution function is non-Maxwellian or Maxwellian.
As a result, the system can hardly maintain stability via external power input less than the fusion power, which means no net power.

In recent years, advancements in instrumentation have led to the re-evaluation of fusion cross-sectional data, with a more accurate identification of the orbital momentum of the primary $\alpha$ particle as $l = 3$ \cite{Stave2011, Spraker2012, Sikora2016}. As a result, Rider's conclusions may need to be revisited in light of these updated cross-sectional data.

In this paper, we conducted theoretical analyses and a series of simulations based on Rider's findings and incorporating the new cross sections. 
We have obtained the temporal dependent recirculating power from simulations, where the recirculating power gradually abates throughout the transition from a specified non-thermal equilibrium state towards a state of thermal equilibrium, considering the adjustment of the Coulomb logarithm's effects on collisional interactions within the context of a particular density. 
Additionally, the impact of ion distribution on the progression of the electron's non-Maxwellian distribution function has been examined.
Under Rider's parameters with ion temperature $T_i$ at 300 keV and mean electron energy $\langle E_e \rangle$ at 35 keV, the associated Bremsstrahlung radiation loss power is constrained to half of the fusion power, but the recirculating power soars to tens of times the fusion power.
Sustaining $T_i$ at 300 keV and a Coulomb logarithm of 15, while raising $T_e$ above Rider's 23.33 keV limit, will reduce electron-ion energy exchange and increase Bremsstrahlung radiation.
Employing updated p-$^{11}$B fusion cross-sectional data, as $T_e$ near 140 keV, the Bremsstrahlung loss power equals the electron-ion interaction power but is still less than the fusion power. This leads to a net energy yield, corroborating the findings in Putvinski's 2019 work \cite{Putvinski2019}.

The structure of this paper is outlined as follows:
In Section II, we provide a brief introduction to Rider's theoretical model.
Section III presents simulation results under two scenarios: Part a considers a Maxwellian velocity distribution function for ions and a non-Maxwellian distribution for electrons, while Part b examines both ions and electrons with Maxwellian velocity distribution functions.
Section IV analyzes the conditions for the system to achieve a net energy gain when the velocity distributions of both ions and electrons are Maxwellian.
Section V offers an analysis of existing proton-boron fusion schemes.
Finally, discussions and conclusions are presented in Section VI.

\section{Theoretical analysis}
The scenario considered by Rider is spatially uniform and isotropic, therefore the evolution of the particle distribution function in plasma follows the equation \cite{Rider1995phd},
\begin{equation}
\label{evolution-f}
\frac{\partial f_\alpha}{\partial t} =\left ( \frac{\partial f_\alpha}{\partial t} \right )_c,
\end{equation}
where the subscript indicates the particle species, and $f_\alpha(\boldsymbol{r},\boldsymbol{v},t)$ represents the distribution function of $\alpha$ particles. 
The collision operator can be written as
\begin{eqnarray}
\label{c operater}
\left ( \frac{\partial f_\alpha}{\partial t} \right )_c &=& \sum_\beta \Gamma_{\alpha \beta} \nabla_{\boldsymbol{v}} \cdot [\frac{\nabla_{\boldsymbol{v}}}{2} \cdot (f_\alpha \nabla_{\boldsymbol{v}} \nabla_{\boldsymbol{v}} g_{\alpha \beta}) - f_\alpha \nabla_{\boldsymbol{v}} h_{\alpha \beta}] \nonumber \\
&\equiv& \sum_\beta C_{\alpha \beta} \equiv -\nabla_{\boldsymbol{v}} \cdot \sum_\beta \boldsymbol{J}_{\alpha \beta},
\end{eqnarray}
where $h_{\alpha \beta}$ and $g_{\alpha \beta}$ represent the Rosenbluth potentials,
\begin{equation}
	\label{Pi_ab}
	\Gamma_{\alpha \beta} = \frac{4\pi Z_\alpha^2 Z_\beta^2 \ln \Lambda}{m_\alpha^2},
\end{equation}
$C_{\alpha \beta}$ denotes the collision operator between $\alpha$ and $\beta$ particles, $\boldsymbol{J}_{\alpha \beta}$ represents the particle flux in velocity space, and the summation symbol encompasses all particle species, including the case where $\beta=\alpha$.
Therefore, the energy transfer rate per volume from the $\alpha$ species to the $\beta$ species is defined as $P_{\alpha \beta}$:
\begin{equation}
	\label{P_ab}
	P_{\alpha \beta} = -\int d^3 \boldsymbol{v} \left ( \frac{1}{2}m_\alpha v^2 \right ) C_{\alpha \beta}.
\end{equation}
Rider provided a comprehensive isotropic particle velocity distribution $f(v)$ (for $v\ge0$), peaking at a velocity $v_0$, with characteristic widths $v_{ts}$ and $v_{tf}$ on the slow and fast sides of the peak, respectively:
\begin{widetext}
	\begin{equation}
	f(v) \equiv \left\{
	\begin{aligned}
	&nK \left\{  \exp\left[-(v-v_0)^2/v_{ts}^2\right]   +  \exp\left[-(v+v_0)^2/v_{ts}^2\right]  \right\}  \ \mathrm{for}\ v < v_0, \\
    &nK \left\{  \exp\left[-(v-v_0)^2/v_{tf}^2\right]   +  \exp\left[-(v+v_0)^2/v_{ts}^2\right]  \right\}  \ \mathrm{for}\ v \ge v_0.
	\end{aligned}
	\right. \label{dist}
	\end{equation}
\end{widetext}
The distribution function of electrons is expressed using $f(v)$ above, which can be both non-Maxwellian and Maxwellian. In Rider's investigation, the ion distribution function is constrained to a Maxwellian distribution. We adopt the same approach.
\hypertarget{fig1}{}
\begin{figure*}[htbp]
	\centering
	\includegraphics[width=0.55\textwidth]{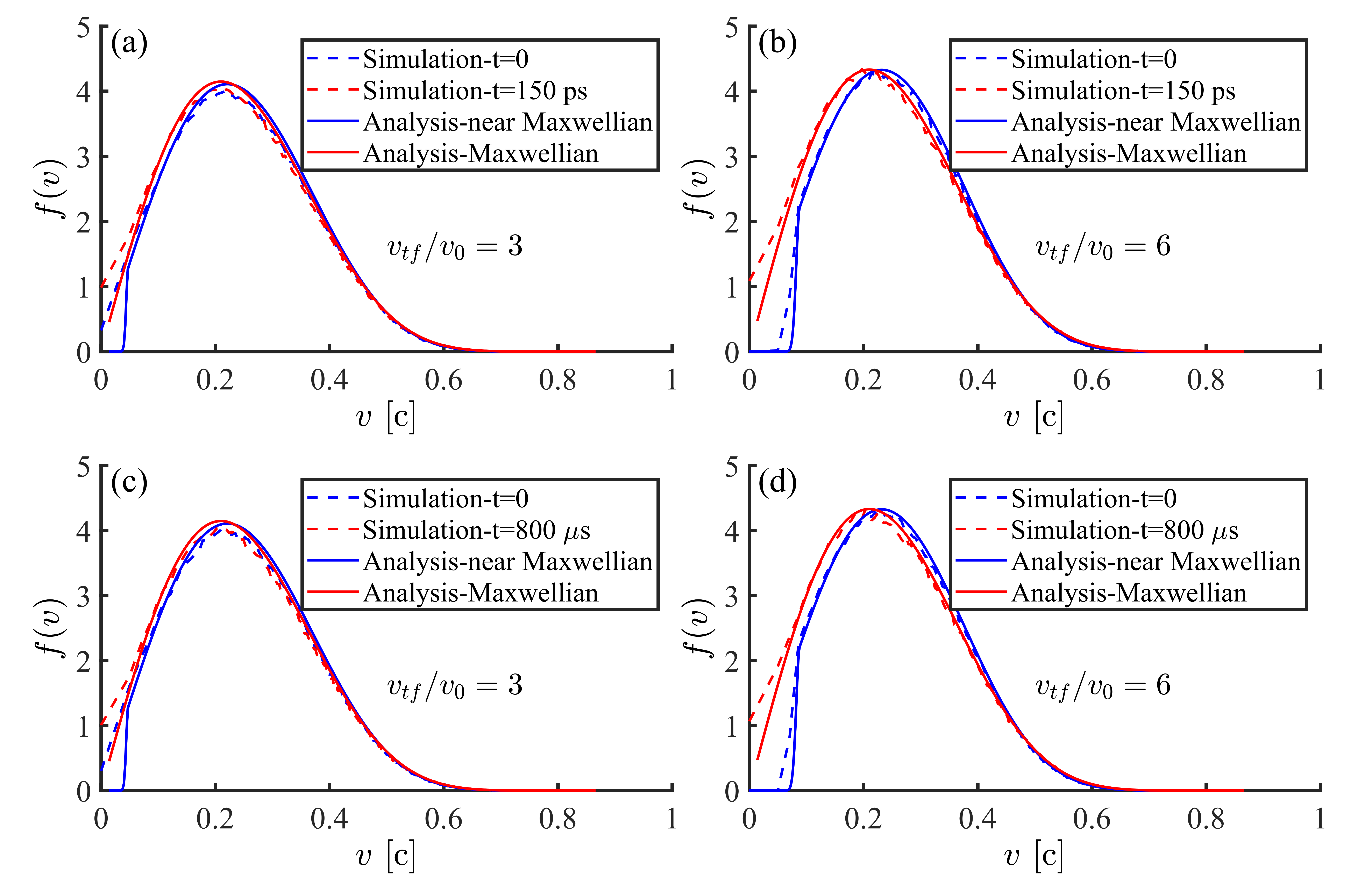}
	\caption{The normalized electron distribution function $f(v)$ is shown between the initial moment and when it approaches thermal equilibrium. The ratio of $v_0/v_{ts}$ is fixed at 10, and $\langle E_e \rangle$ is held constant at 35 keV. In (a)-(b), the number density of electrons $n_e$ is $1\times 10^{28}\ \mathrm{m}^{-3}$, while in (c)-(d), $n_e$ is $1\times 10^{20}\ \mathrm{m}^{-3}$.}
	\label{Fig:1}
\end{figure*}

\section{Fusion System in Non-thermal Equilibrium State}
The fusion system considered by Rider exits in a non-thermal equilibrium state. 
Rider explored two scenarios of non-equilibrium states \cite{Rider1997}: one is that the electron distribution function derives from Maxwellian, and the ion temperature $T_i$ surpasses the effective electron temperature $T_\mathrm{eff}=2\langle E_e \rangle /3$, where $\langle E_e \rangle$ represents the mean energy of electrons.
The other scenario involves electrons being in their own thermal equilibrium, while the ion temperature $T_i$ exceeds the electron temperature $T_e$, akin to the commonly termed hot ion mode.

In these circumstances, a low average electron energy results in restricted Bremsstrahlung loss, whereas high ion temperature favors fusion processes. 
Nonetheless, the non-thermal equilibrium state cannot persist spontaneously, as the system has not attained its ultimate thermal equilibrium.
Energy transfer serves as the primary mechanism for achieving thermodynamics equilibrium and must be hindered to keep the fusion system stable. The rate of energy exchange between species $\alpha$ and $\beta$ (including $\alpha = \beta$) can be calculated using Eq.\ (\ref{P_ab}).

Meanwhile, the fusion fuel ions undergo nuclear reactions
\begin{equation}
	p\ +\ ^{11}B=3\alpha + 8.7\ \mathrm{MeV}.
\end{equation}
Generally speaking, the fusion power per volume is given by
\begin{equation}
	P_{fus} = n_1 n_2 \langle \sigma v\rangle Q_{\mathrm{fus}},\label{fus}
\end{equation}
where $n_1,\ n_2$ represent the number densities of the two reactants, with the first ion species being proton and the second ion species being boron.
$\langle \sigma v\rangle$ represents the average fusion reactivity, and $Q_{\mathrm{fus}}$ is the energy released per reaction, for p-$^{11}$B fusion $Q_{\mathrm{fus}}=8.68$ MeV.

Bremsstrahlung radiation is taken into account for assessing the performance of fusion systems.\ The Bremsstrahlung power loss per volume, including relativistic corrections, is provided by Rider in 1995 \cite{Rider1995}, which aligns well with H. S. Xie's fitting formula \cite{Xie2024}. 
Xie observed that Rider's total fitting approach, although it tends to overestimate the ion contributions and underestimate the electron contributions, yields a result that closely aligns with the exact value.

\subsection{Non-Maxwellian Electron Distribution}
In the case where the electrons exhibit a near-Maxwellian distribution with the slow electrons depleted ($\langle E_i \rangle \gg \langle E_{ie} \rangle$), the energy transfer between ions and electrons primarily involves low energy electrons \cite{Rider1995phd}, 
constrained to a minimal level.\ During such instances, the energy transfer predominately transpires among electrons. 
\hypertarget{fig2}{}
\begin{figure}[htbp]
	\centering
	\includegraphics[width=0.35\textwidth]{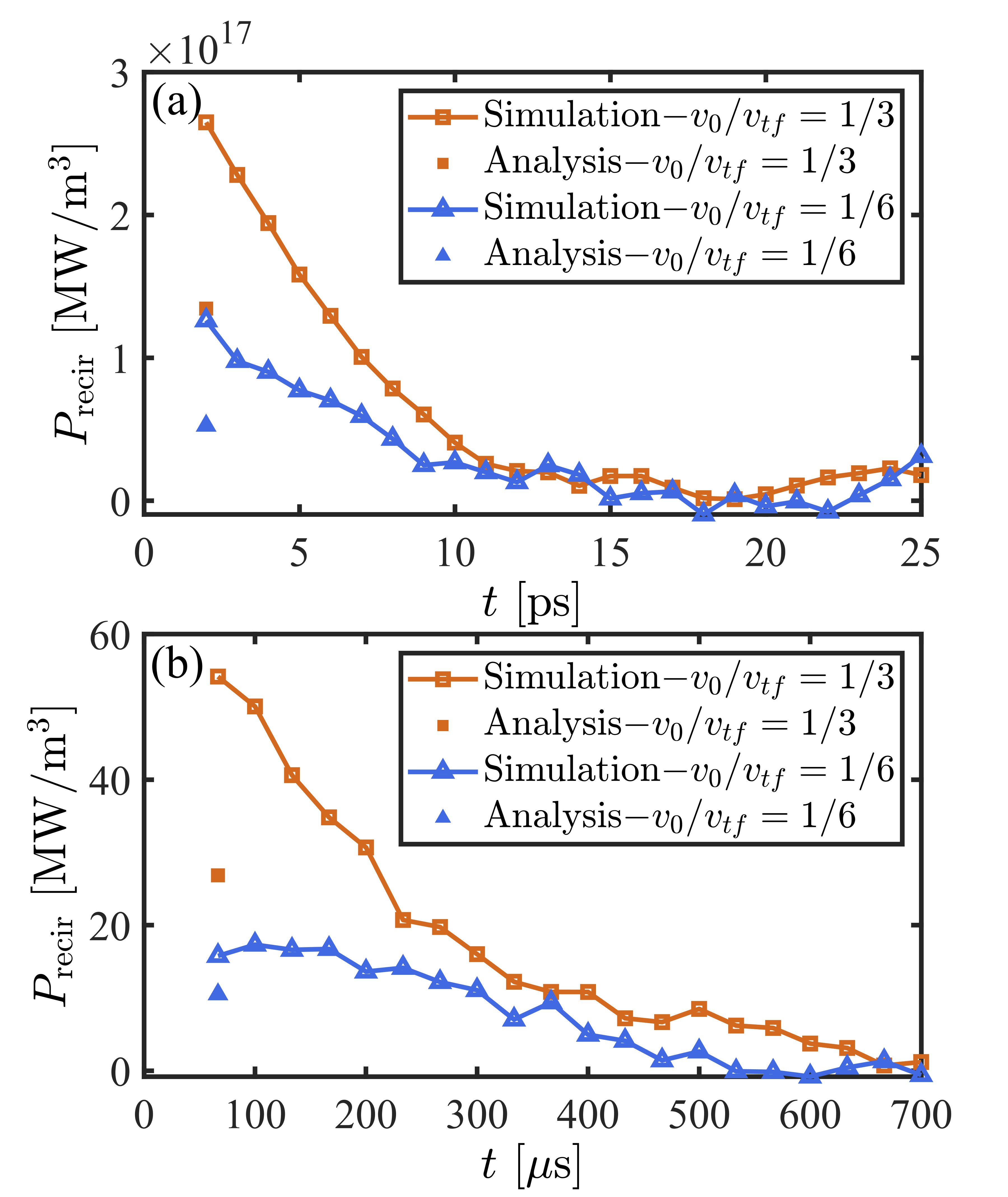}
	\caption{The required recirculating power to preserve the nearly Maxwellian distribution of electrons as a function of time. The electron density $n_e$ is (a) $1\times 10^{28}\ \mathrm{m^{-3}}$ and (b) $1\times 10^{20}\ \mathrm{m^{-3}}$, respectively. When $v_0/v_{tf}=1/6$, $v_{tf} = 8.5\times10^7\ \mathrm{m/s}$, and when $v_0/v_{tf}=1/3$, $v_{tf} =7.95\times10^7\ \mathrm{m/s}$.}
\end{figure}

\hypertarget{fig3}{}
\begin{figure*}[htbp]
	\centering
	\includegraphics[width=0.7\textwidth]{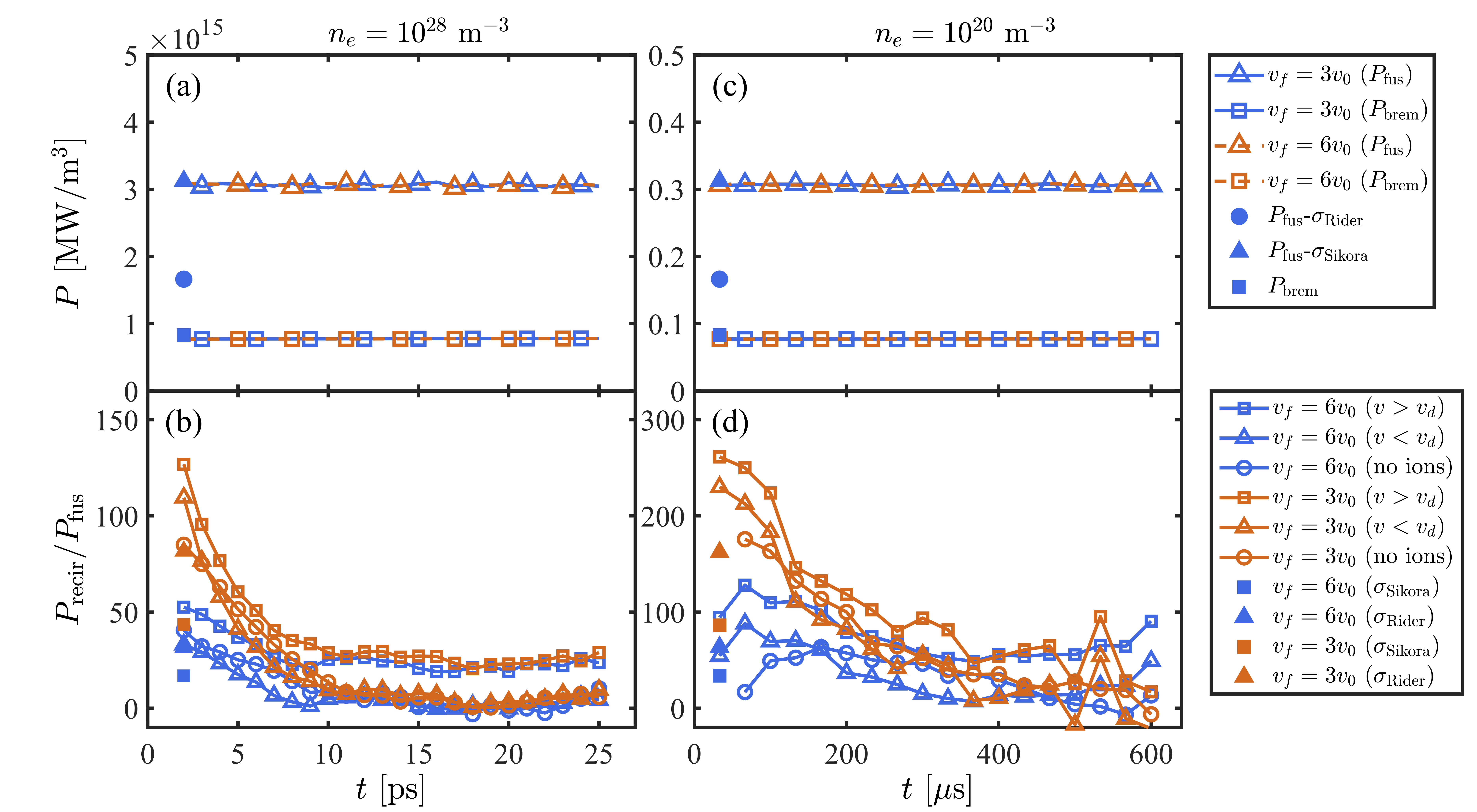}
	\caption{(a) and (c) depict the plots of $P_\mathrm{fus}$ and $P_\mathrm{brem}$ as functions of simulated time for electron densities $n_e$ of $10^{28}\ \mathrm{m}^{-3}$ and $10^{20}\ \mathrm{m}^{-3}$, respectively. The solid markers represent theoretical results, while line markers denote simulation results. (b) and (d) illustrate the plots of ratio $P_\mathrm{recir}/P_\mathrm{fus}$ against simulated time for $n_e$ of $10^{28}\ \mathrm{m}^{-3}$ and $10^{20}\ \mathrm{m}^{-3}$, respectively.}
	\label{Fig:3}
\end{figure*}
A certain portion of electrons ($N_\mathrm{slow}$) have slowed down due to energy loss from collisions, while others ($N_\mathrm{fast}$) have accelerated for the same reason and become too fast. 
Sustaining the non-Maxwellian electron distribution requires adding a specific amount of energy to the slow electrons whenever they experience energy depletion, while withdrawing an equivalent amount of energy from the fast electrons. Ideally, the energy supplied to the slow electrons could be derived by selectively extracting energy from the fast ones. 

The rate of energy transfer between the fast electrons and the slow electrons is given by \cite{Rider1997}
\begin{eqnarray}
	\label{P_ee}
	P_{ee}&=&-\int_0^{v_d} (dv\ 4\pi v^2) \left (\frac{1}{2}mv^2 \right)\left ( \frac{\partial f}{\partial t} \right ) _{\mathrm{col}} \nonumber \\
	&=&\int_{v_d}^{\infty} (dv\ 4\pi v^2) \left (\frac{1}{2}mv^2 \right)\left ( \frac{\partial f}{\partial t} \right ) _{\mathrm{col}},
\end{eqnarray}
where $v_d$ denotes the dividing velocity to distinguish the slow electrons and fast electrons, and the $(\partial f/\partial t)_\mathrm{col}$ represents the collision operator acting between the electrons.\ The recirculating power $P_\mathrm{recir}$ that must be recycled to maintain the non-Maxwellian electron distribution is consequently equivalent to $P_{ee}$, which is proportional to $n_e^2 \mathrm{ln} \Lambda$.

We conducted a series of simulations using {\small LAPINS} code \cite{DWu2017, DWu2018, DWu2019, DWu2021} based on the fusion system outlined by Rider.\
Theoretically, $P_{ie}$ is assumed to be a small quantity under this situation. 
We initially simulate the case considering only electrons. The velocity distribution of the electrons employed in the {\small LAPINS} code is described by Eq.\ (\ref{dist}).
Fig.\ \hyperlink{fig1}{1} illustrates the normalized electron distribution function $f(v)$ at the initial moment and as it approaches thermal equilibrium.

By calculating Eq.\ (\ref{P_ee}), we can determine the required recirculating power $P_{\mathrm{recir}}$ to maintain a nearly Maxwellian distribution of electrons. 
Throughout the simulation, the electron energy spectrum at various time points can be obtained. Calculating $v_d$ and submitting it into Eq.\ (\ref{P_ee}), the rate of energy transfer between slow and fast electrons can be obtained, which is equivalent to $P_\mathrm{recir}$.
The outcomes calculated for $v>v_d$ coincides with those calculated for $v<v_d$, which aligns with Rider's theoretical predictions. Within the code, the ratio $v_0/v_{ts}$ remains fixed at 10, while the average electron energy $\langle E_e\rangle$ is held constant at 35 keV. As depicted in Fig.\ \hyperlink{fig2}{2}, for the same $\langle E_e\rangle$, higher values of $v_{tf}/v_0$ lead to increased recirculating power $P_\mathrm{recir}$ due to varying degrees of deviation from the Maxwellian distribution.
\hypertarget{fig4}{}
\begin{figure}[htbp]
	\centering
	\includegraphics[width=0.35\textwidth]{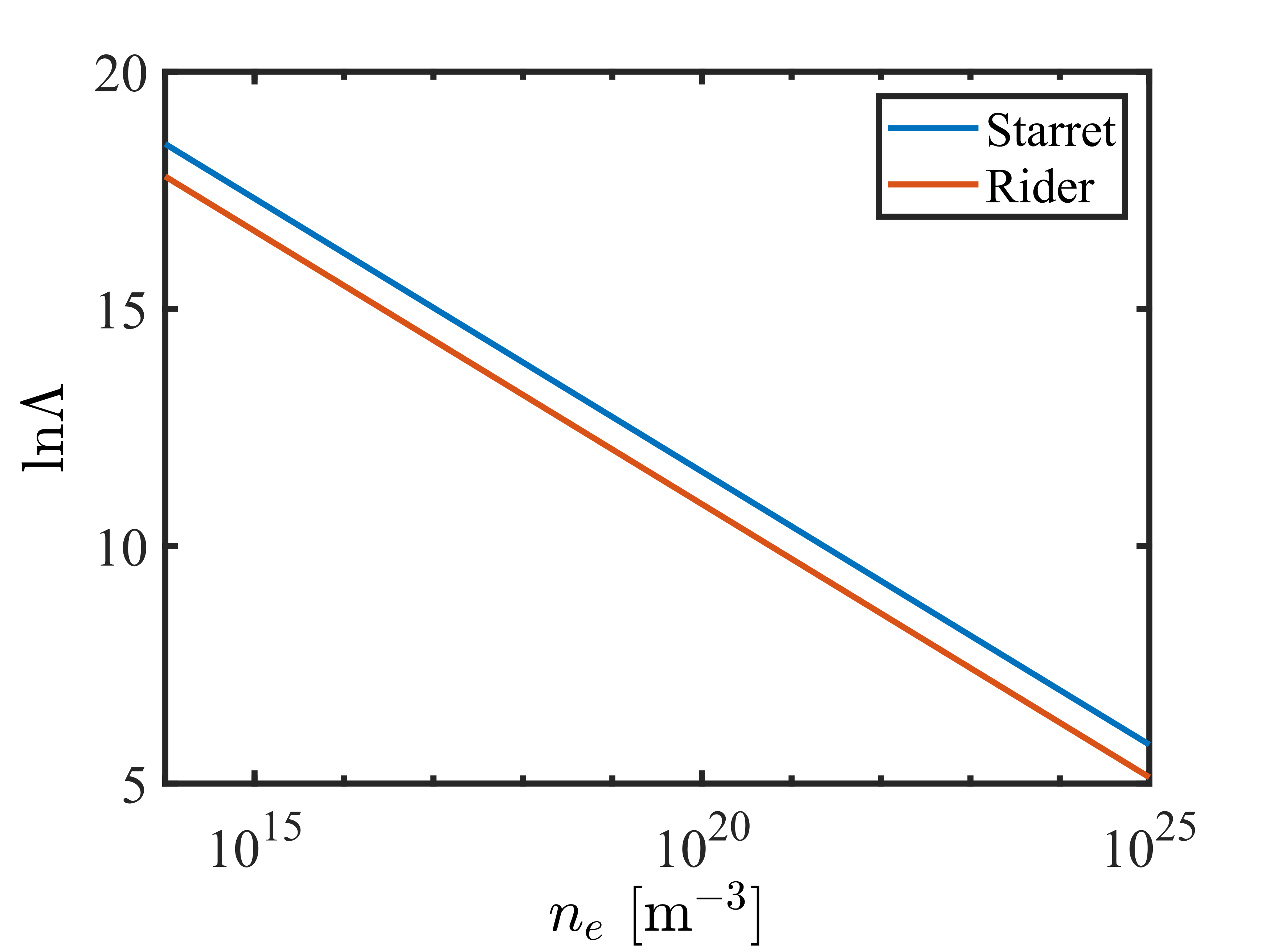}
	\caption{Coulomb logarithm $\mathrm{ln}\Lambda$ is characterized as a function of electron density. $T_i=300$ keV, $T_\mathrm{eff}=35$ keV, and $n_p=5n_B$. The blue line is plotted from Starret'work \cite{Starrett2018}, and the red line is plotted from Rider's empirical formula.}
	\label{Fig:4}
\end{figure}

In practical scenarios, the impact of ions on the evolution of the electron distribution function cannot be overlooked. Therefore, we simulate the situation where both ions and electrons are considered, and nuclear reactions and Bremsstrahlung processes are considered together.
\hypertarget{fig5}{}
\begin{figure*}[!htbp]
	\centering
	\includegraphics[width=0.7\textwidth]{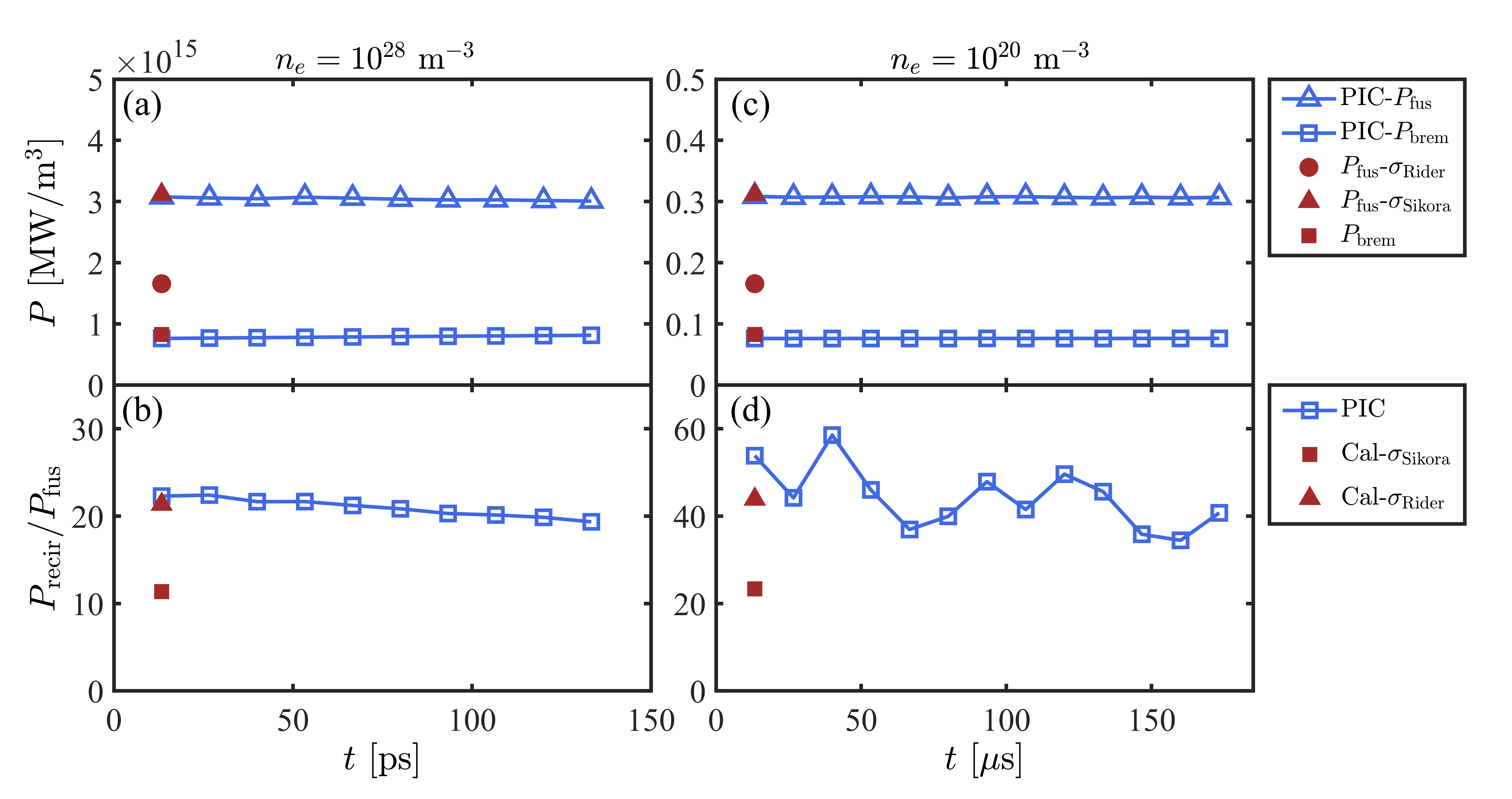}
	\caption{(a) and (c) depict the plots of fusion power and Bremsstrahlung power loss as functions of simulated time for electron densities $n_e$ of $10^{28}\ \mathrm{m}^{-3}$ and $10^{20}\ \mathrm{m}^{-3}$, respectively. The lines represent theoretical results, while line markers denote simulation results. (b) and (d) illustrate the plots of recirculation power to fusion power ratio $P_\mathrm{recir}/P_\mathrm{fus}$ against simulated time for electron densities $n_e$ of $10^{28}\ \mathrm{m}^{-3}$ and $10^{20}\ \mathrm{m}^{-3}$, respectively.}
	\label{Fig:5}
\end{figure*}

The simulation results regarding the recirculating power $P_{\mathrm{recir}}$, the fusion power $P_\mathrm{fus}$ and the Bremsstrahlung power $P_\mathrm{brem}$ are depicted in Fig.\ \hyperlink{fig3}{3}. 
As illustrated in Fig.\ \hyperlink{fig3}{3} (b) and (d), $P_{\mathrm{recir}}$ exhibits similar trends regardless of the inclusion of ions. However, when the fusion ions are included, the result obtained from $v>v_d$ calculations is marginally higher than that from $v<v_d$ calculations.\
This discrepancy arises due to the additional energy boost imparted to high-energy electrons by ion-electron and electron-electron collisions.

In Fig.\ \hyperlink{fig3}{3} (a), the fusion power and Bremsstrahlung power calculated demonstrate good agreement with simulation results. 
The cross section for p-$^{11}$B fusion utilized by Rider is based on McNally's research from 1979 \cite{Mcnally1979}, which yields a value approximately half of that reported by Sikora in 2016 \cite{Spraker2012, Sikora2016}.
Consequently, the fusion power calculated using the updated cross section is approximately double that computed by Rider.
Moreover, about the effect of density on the ratio $P_{\mathrm{recir}}/P_{\mathrm{fus}}$, it can be derived that $P_{\mathrm{recir}}/P_{\mathrm{fus}} \propto n_e^2\mathrm{ln}\Lambda/n_1n_2\propto\mathrm{ln}\Lambda$. 
As shown in Fig.\ \hyperlink{fig4}{4}, the lower density results in a higher $\mathrm{ln}\Lambda$, therefore the ratio $P_{\mathrm{recir}}/P_{\mathrm{fus}}$ is higher at $n_e=10^{20} \ \mathrm{m}^{-1/3}$ compared to $n_e=10^{28}\ \mathrm{m}^{-1/3}$.

\subsection{Maxwellian Electron Distribution}
When both the ions and electrons adhere to a Maxwellian distribution, the primary mechanism for dissipating the non-equilibrium state is through energy transfer between ions and electrons. The energy transfer rate $P_{ie}$ can be determined using corrected Spitzer's formula \cite{Rider1995}. In this case, the energy source of electrons originates from $P_{ie}$, while energy loss stems from $P_\mathrm{brem}$, therefore the minimum recirculating power is $P_\mathrm{recir}\equiv P_{ie}-P_\mathrm{brem}$. 

We also provide data on fusion power and Bremsstrahlung power, illustrated in Fig.\ \hyperlink{fig5}{5}(a) and (c).\
Similarly, we depict the ratio of recirculating power to fusion power in Fig.\ \hyperlink{fig5}{5}(b) and (d). 
Upon contrasting the results of electron distribution functions between Maxwellian and non-Maxwellian distributions, it is observed that the recirculating power is slightly lower when the electron distribution function follows a Maxwellian distribution.\
Furthermore, owing to discrepancies between the collision models utilized in the simulation code and theoretical forecasts, the simulated outcomes are roughly double those of theoretical predictions. 
In addition, the effect of $\mathrm{ln}\Lambda$ results in a higher $P_\mathrm{recir}/P_\mathrm{fus}$ ratio at lower densities compared to higher densities. However, $P_\mathrm{brem}/P_\mathrm{fus}$ is independent of density, so the ratio remains constant as the density of electrons changes. Detailed results are shown in Fig.\ \hyperlink{fig6}{6}.
\hypertarget{fig6}{}
\begin{figure}[htbp]
	\centering
	\includegraphics[width=0.45\textwidth]{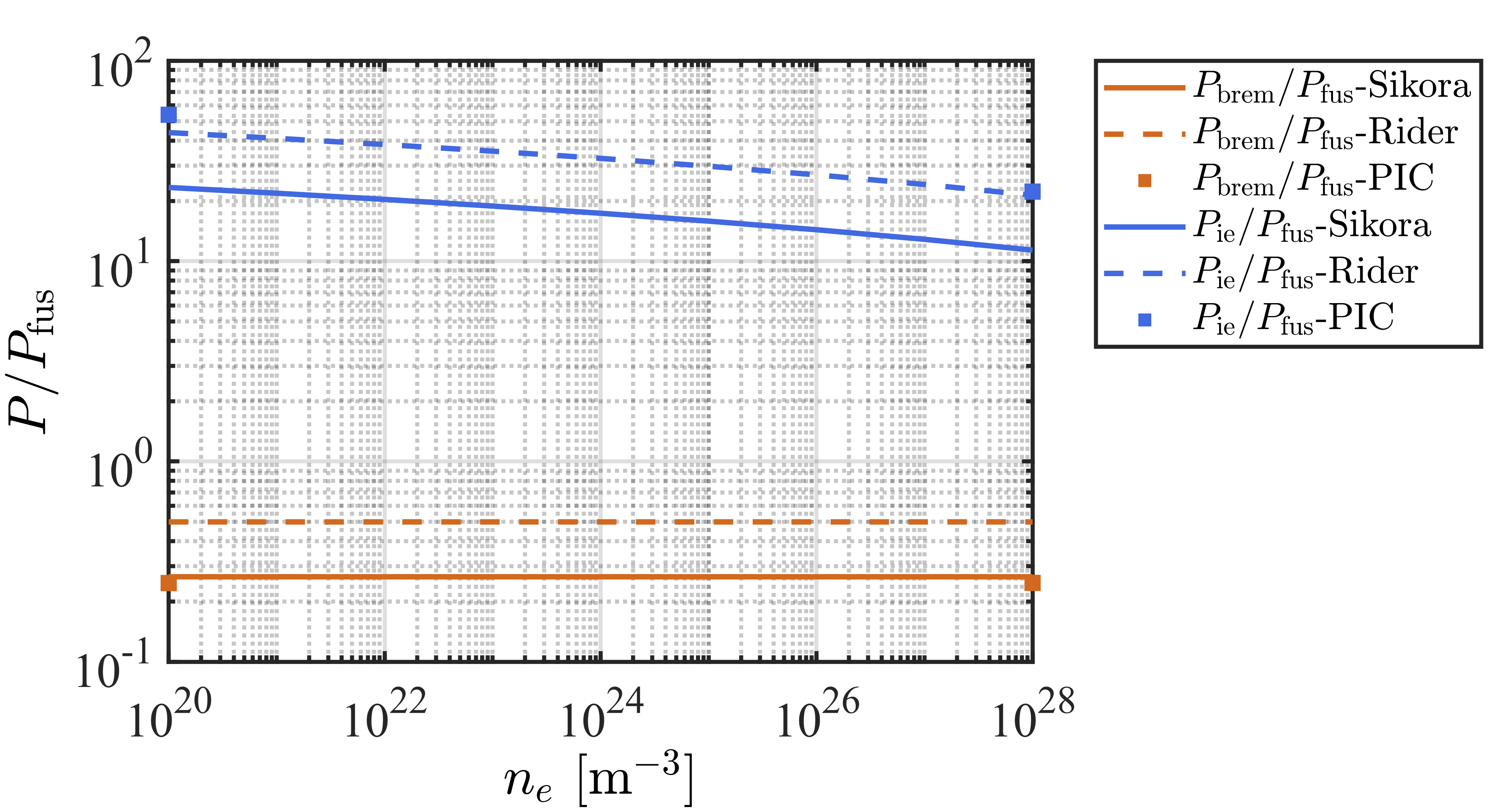}
	\caption{The ratio of Bremsstrahlung power and the energy transfer rate between ions and electrons to fusion power as functions of electron densities $n_e$.}
	\label{Fig:6}
\end{figure}

\section{the conditions for the system to achieve a net energy gain}
Based on the preceding analysis, we are inclined to approach the consideration of energy gain in non-thermal equilibrium fusion systems starting from the assumption that the electron distribution function is Maxwellian.
Taking into account the work of Putvinski in 2019 \cite{Putvinski2019}, achieving the self-sustainability in the p-$^{11}$B reaction is possible. Following Rider's research, we can delineate the intermediate processes between them. Illustrated in Fig.\ \hyperlink{fig7}{7}(a), with the ion temperature set at 300 keV, $P_\mathrm{ie}/P_\mathrm{fus}$ and $P_\mathrm{brem}/P_\mathrm{fus}$ can be conveniently calculated. 
The dashed line represents calculations using $\left \langle \sigma v\right \rangle$ from Rider's reference, while the solid line employs the new nuclear reaction cross section for 2016 \cite{Sikora2016}. The logarithm is set at 15 in accordance with Rider's study.
In Fig.\ \hyperlink{fig7}{7}(a), the left end of the line corresponds to Rider's parameters. For lower $\left \langle \sigma v\right \rangle$, $P_\mathrm{ie}/P_\mathrm{fus}=36$ and $P_\mathrm{brem}/P_\mathrm{fus}=0.5$. For higher $\left \langle \sigma v\right \rangle$, $P_\mathrm{ie}/P_\mathrm{fus}=19.5$ and $P_\mathrm{brem}/P_\mathrm{fus}=0.27$. 
Considering the system assumed by Rider \cite{Rider1997} and using the lower $\left \langle \sigma v\right \rangle$, to maintain the stability of the p-$^{11}$B plasma fusion systems not in thermodynamic equilibrium, the relationship between $P_\mathrm{recir}$ and $P_\mathrm{fus}$ could be derived,
\begin{eqnarray}
&P_\mathrm{ie} = P_\mathrm{recir}+P_\mathrm{brem}, \nonumber \\
&\eta P_\mathrm{recir}+P_\mathrm{fus}=P_\mathrm{ie}
\end{eqnarray}
where $\eta$ represents the efficiency of the external heat engines, $\eta<1$. Achieving a net energy output requires $\eta$ to exceed the value indicated by the dashed line in Fig.\ \hyperlink{fig7}{7}(b), which is nearly unattainable.\
\hypertarget{fig7}{}
\begin{figure}[htbp]
	\centering
	\includegraphics[width=0.35\textwidth]{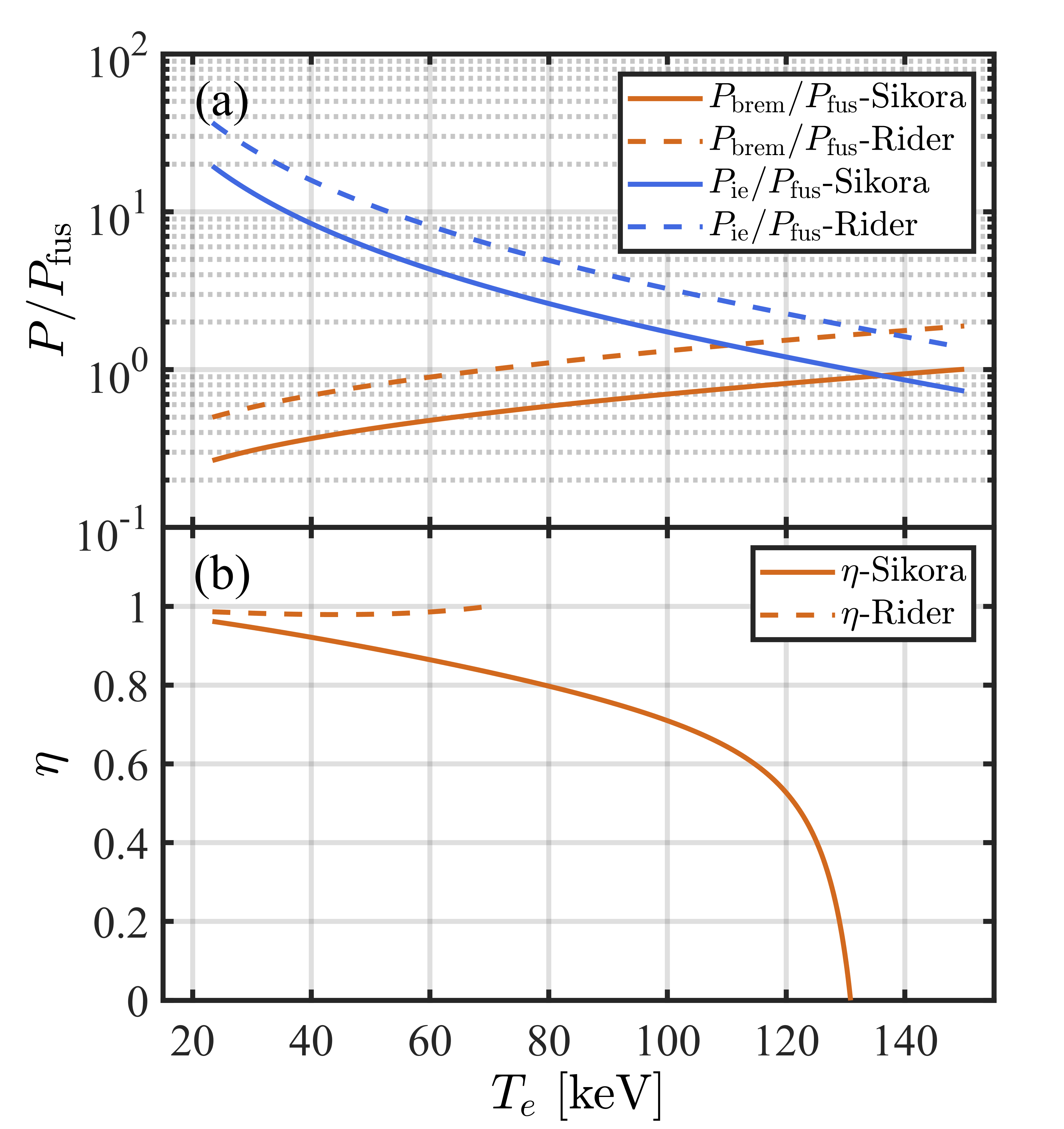}
	\caption{(a) The ratio $P/P_\mathrm{fus}$ as a function of electron temperature. The Coulomb logarithm is set to 15 and $T_i$ is fixed to 300 keV. (b) To maintain the stability of the p-$^{11}$B plasma fusion systems not in thermodynamic equilibrium, the required efficiency $\eta$ on the heat engine as a function of electron temperature.}
	\label{Fig:7}
\end{figure}
However, with a higher $\left \langle \sigma v\right \rangle$, when $T_i$ ranges from about 250 to 350 keV, at this point in thermal equilibrium, $P_\mathrm{fus}$ exceeds $P_\mathrm{brem}$ \cite{Putvinski2019}. 
Setting $T_i$ to 300 keV, when $P_\mathrm{brem}=P_\mathrm{ie}$, $T_e=136.1$ keV, at this time, $P_\mathrm{fus}-P_\mathrm{brem}=8.31\% P_\mathrm{fus}$, indicating a net energy output. 
To obtain a net energy output, $\eta$ must surpass the value represented by the solid line in Fig.\ \hyperlink{fig7}{7}(b).
Owing to the difficulties associated with the recycling of the Bremsstrahlung radiation, we do not take into account the instances where $P_\mathrm{brem}>P_\mathrm{fus}$.
The new nuclear reaction cross-section yields a result that is much more optimistic compared to the cross-section used by Rider, however, it demands high temperatures for both ions and electrons.

Based on the aforementioned analysis, achieving net energy gain from p-$^{11}$B fusion in a non-equilibrium state necessitates recycling the power $P_\mathrm{recir}$ originated from collisions to sustain the non-equilibrium condition. 
External heat engines, utilizing the heat generated during the evolution from non-equilibrium to thermal equilibrium as a high-temperature heat source, theoretically represent the most efficient form of energy extraction to recycle the power. 
However, this method requires a highly efficient external heat engine, or it requires high temperatures for ions and electrons to achieve a net energy gain.
Rider pointed out that in theory, energy can be selectively harnessed from specific regions within the velocity space of particle distributions, and this harvested energy could then, theoretically, be processed and reintroduced into different regions of the particle velocity space \cite{Rider1997}.
Furthermore, if there exists a ``internal-engine" mechanism by which electrons with thermal velocities far greater than those of ions can transfer energy to ions, such as through wave-particle interactions, this could potentially aid in the maintenance of a non-equilibrium state.

\section{Analysis of existing proton-boron fusion schemes}
Currently, several companies are engaged in the research of p-$^{11}$B fusion, such as TAE Technologies \cite{Gota2021}, HB11 \cite{Hora2017}, and ENN \cite{liums2024}. Notably, Energy iNNovation (ENN) Science and Technology Development Co. has planned to achieve hydrogen-boron fusion through a spherical torus design.
In this paper, we conduct a preliminary analysis based on the potential parameters of ENN's future endeavors.

The parameters for achieving p-$^{11}$B fusion gain ($Q>1$) at present can be as follows \cite{liums2024}:\ magnetic field strength ($B_0$) in the range of 4-7 T, ion temperature to electron temperature ratio ($T_i/T_e$) of $\ge3$, ion temperature ($T_{i0}$) between 100-300 keV, electron density ($n_{e0}$) of 1-$3\times10^{20}\ \mathrm{m}^{-3}$, and energy confinement time ($\tau_E$) ranging from 20-50 s. Using the new Spherical Torus energy confinement scaling law, a reactor with major radius ($R_0$) of 4 m, central magnetic field ($B_0$) of 6 T, central temperature ($T_{i0}$) of 150 keV, plasma current ($I_p$) of 30 MA, and hot ion mode ($T_i/T_e$) of 4 can achieve p-$^{11}$B fusion with $Q>10$.

\hypertarget{fig8}{}
\begin{figure}[htbp]
	\centering
	\includegraphics[width=0.35\textwidth]{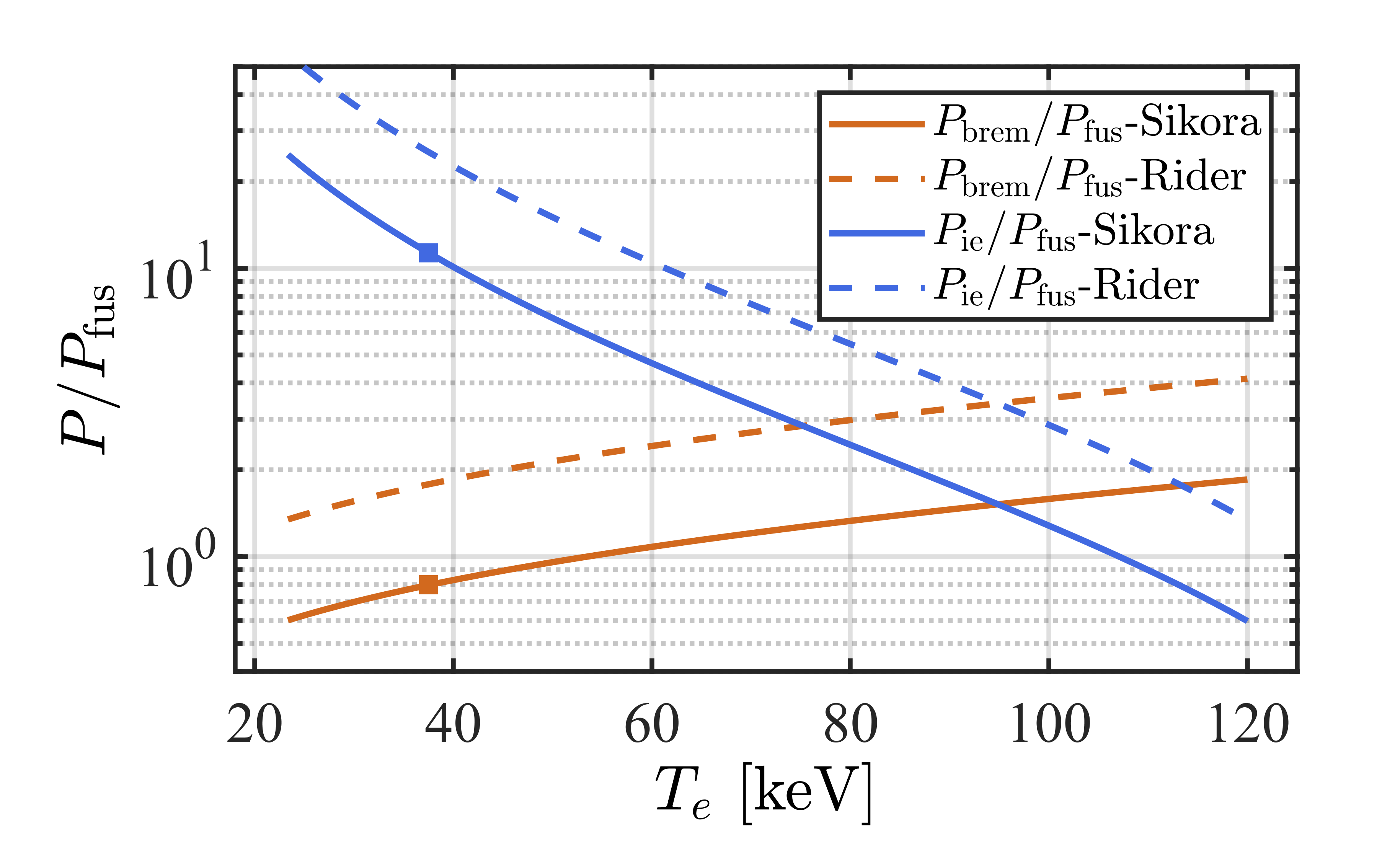}
	\caption{The ratio $P/P_\mathrm{fus}$ as a function of electron temperature. The Coulomb logarithm is set to 18 and $T_i$ is fixed to 150 keV.}
	\label{Fig:8}
\end{figure}

This paper does not involve specific configurations or magnetic fields, and thus we have not considered synchrotron radiation in our calculations.
However, it should be noted that in magnetic confinement scenarios, unless the synchrotron radiation is almost entirely reflected, it is necessary to account for corrections due to synchrotron radiation \cite{Cai2022}.
Based on the data from the ENN, Fig.\ \hyperlink{fig8}{8} illustrates the variation of the ratios $P_\mathrm{brem}/P_\mathrm{fus}$ and $P_\mathrm{ie}/P_\mathrm{fus}$ with respect to electron temperature.
Specifically, for a scenario where $T_i=150$ keV, $T_i/T_e=4$, electron density $n_{e}=1$-$3\times10^{20}\ \mathrm{m}^{-3}$ and $n_p=5 n_B$, with a corresponding Coulomb logarithm of approximately 18, the obtained results show $P_\mathrm{brem}/P_\mathrm{fus} = 0.80$, $P_\mathrm{ie}/P_\mathrm{fus}=11$. Notably, the substantial rate of ion-electron energy transfer significantly complicates achieving a favorable net energy output based on p-$^{11}$B fusion.

\section{Discussions and Conclusions}
Rider made several assumptions in his work \cite{Rider1997}.\
In our study, we largely adhered to Rider's assumptions, with a few additional considerations:\
1.\ Concerning the energy sources available to electrons, the {\small LAPINS} code incorporates interactions between fusion ions and fusion-generated ions with electrons. 2.\ We considered the dependency of the Coulomb logarithm on density, particularly evident in higher Coulomb logarithms at lower densities, leading to a corrective effect on the recirculating power.

Additionally, under specific conditions where ions follow a non-Maxwellian distribution, it is possible to induce fusion reaction rates that exceed those anticipated by Maxwellian distribution \cite{Xie2023}.\
However, the minimum recirculating power required to recycled to sustain such non-Maxwellian distributions necessitates detailed assessment.
Furthermore, the assumption by Rider that ``the Bremsstrahlung power is approximately independent of the electron velocity distribution shape in the energy range of interest" needs to be reconsidered, especially when the effective electron temperature exceeds 100 keV.
This consideration should not be overlooked in determining whether a net energy gain can be achieved in the p-$^{11}$B reaction under practical conditions.

In this study, we have revisited Rider's investigation concerning the fundamental power constraints inherent in p-$^{11}$B fusion reactors, particularly when the electrons exhibit a non-Maxwellian velocity distribution or when electrons and ions are characterized by substantially divergent temperatures.
Theoretical computations grounded on Rider's framework, alongside simulations conducted utilizing the {\small LAPINS} code, have been executed to ascertain the fusion power, Bremsstrahlung power, and recirculating power.
While the updated cross section for p-$^{11}$B fusion promises enhanced fusion power and the {\small LAPINS} code adopts a distinct approach in addressing collision effects, the magnitude of recirculating power that must be recycled to maintain the plasma in a state of non-equilibrium significantly surpasses the fusion power output. 
Consequently, reactors utilizing inherently non-equilibrium plasmas for p-$^{11}$B fusion will prove incapable of generating net power unless equipped with exceedingly efficient external heat engines capable of recycling the recirculating power.

However, the non-thermal equilibrium states considered by Rider deviate significantly from thermal equilibrium. Maintaining the ion temperature constant, an elevation of the electron temperature beyond Rider's consideration of 23.33 keV results in a decrease in the electron-ion energy transfer rate, coupled with an increase in Bremsstrahlung power. When the electron temperature reaches approximately 140 keV, fusion power exceeds both Bremsstrahlung power and electron-ion energy transfer rate, indicating a net energy output.

\section{Acknowledgements}
This work was supported by Energy iNNovation (ENN) Science and Technology Development Co. (Grant No. 2021HBQZYCSB006), the National Natural Science Foundation of China (Grant No. 11875235). The discussions with Dr. Hua-sheng Xie are acknowledged.

\bibliography{reference}
\end{document}